\documentclass[aps,prl,floatfix,twocolumn,superscriptaddress,twoside]{revtex4-1}
\usepackage{amsmath,amssymb,amsfonts}
\usepackage{graphicx}
\usepackage[english]{babel}
\usepackage{psfrag}
\usepackage{color}
\usepackage{tabularx}
\usepackage{times}
\usepackage{hyperref}
\usepackage{float}
\usepackage{gensymb}
\begin{document}

\title{Quantum spin liquid phase in the Shastry-Sutherland model detected by an \\ improved level spectroscopic method}

\author{Ling Wang}
\email{lingwangqs@zju.edu.cn}
\affiliation{Department of Physics, Zhejiang University, Hangzhou 310000, China}

\author{Yalei Zhang}
\affiliation{Beijing Computational Science Research Center, 10 East Xibeiwang Road, Beijing 100193, China}

\author{Anders W. Sandvik}
\email{sandvik@bu.edu}
\affiliation{Department of Physics, Boston University, 590 Commonwealth Avenue, Boston, Massachusetts 02215, USA}
\affiliation{Beijing National Laboratory for Condensed Matter Physics and Institute of Physics, Chinese Academy of Sciences, Beijing 100190, China}

\begin{abstract}
We study the spin-$1/2$ two-dimensional Shastry-Sutherland spin model
by exact diagonalization of clusters with periodic boundary conditions. 
We develop an improved level spectroscopic technique using energy gaps 
between states with different quantum numbers. The crossing points of some of 
the relative (composite) gaps have much weaker finite-size drifts than the normally used gaps 
defined only with respect to the ground state, thus allowing precise determination 
of quantum critical points even with small clusters. Our results support the picture
of a spin liquid phase intervening between the well known
plaquette-singlet and antiferromagnetic ground states, with phase
boundaries in almost perfect agreement with a recent density matrix
renormalization group study, where much larger cylindrical lattices
were used [J. Yang et al., Phys. Rev. B {\bf 105}, L060409 (2022)]. The
method of using composite low-energy gaps to reduce scaling corrections 
has potentially broad applications in numerical studies of 
quantum critical phenomena.
\end{abstract}

\date{\today}
\maketitle

{\bf Introduction.}---Quantum spin liquids (QSLs)~\cite{Balents05} are
some of the most intriguing phases of two-dimensional (2D) quantum
matter, yet they have been experimentally elusive. The kagome Heisenberg
antiferromagnet and the Kitaev honeycomb model are among
the most well studied examples. The former hosts a QSL ground state
whose nature was debated for years \cite{White11} but now is largely
settled as a gapless variant \cite{Liao17,He17}. The latter has
exactly solvable gapped and gapless QSL phases \cite{Kitaev06}. Both
models have attracted enormous attention because of their possible
experimental realizations in layered quantum magnets
\cite{Nocera05,Lee07,Khaliullin10}. Recent experiments support gapless
QSLs in both kagome \cite{Norman16,Khuntia20} and honeycomb systems 
\cite{Zheng17,Li21}. Here it should be noted that various defects and
disorder can drastically influence gapless excitations and drive quantum
magnets to randomness-dominated quantum states completely different from the conjectured
pristine gapless QSLs \cite{Kimchi18a,Liu18,Kawamura19}. Experimentally, it is often
difficult to distinguish between these states, as exemplified by contradictory
studies of triangular-lattice systems \cite{Li15,Ma15,Kimchi18b}

\begin{figure}[b]
\includegraphics[width=65mm]{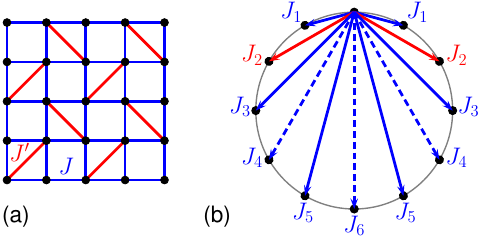}
\vskip-1mm
\caption{Illustration of the spin-$1/2$ models studied here. (a) The SSM, with blue and red lines indicating the 
  Heisenberg AFM interactions $J$ and $J'$, respectively. (b) Heisenberg spin chain with long-range interactions,
  with all couplings of one spin (top of the ring) to all the other spins marked according to the type of coupling.
  The  blue solid and dashed lines mark unfrustrated AFM (odd distances) and ferromagnetic (even distances) couplings,
  respectively, and the red lines show the frustrated AFM $J_2$ interactions.}
\label{hh}
\end{figure}

Another prominent quasi-2D frustrated quantum magnet is
SrCu$_2$(BO$_3$)$_2$ (SCBO)
\cite{Kageyama99,Waki07,Radtke15,Haravifard16,Zayed17,Bettler20,Guo20,Jimenez20,Bettler20},
whose in-plane copper magnetic exchange and super-exchange integrals realize
the inter-dimer ($J$) and intra-dimer ($J^{\prime}$) interactions of
the spin-$1/2$ Shastry-Sutherland model (SSM) \cite{Shastry81}, illustrated in Fig.~\ref{hh}(a). 
Under increasing hydrostatic pressure, the ratio $g\equiv J/J^{\prime}$
increases, and the material undergoes transitions among the three well established 
ground state phases of the SSM; the dimer singlet (DS) phase, a plaquette-singlet 
solid (PSS) phase, as well as an antiferromagnetic (AFM) phase
\cite{Kageyama99,Waki07,Radtke15,Haravifard16,Zayed17,Guo20,Jimenez20,Bettler20}.

Until recently, SCBO was not widely considered as a candidate for a QSL
phase; instead the putative deconfined quantum critical point (DQCP)
separating the PSS and AFM phases was the focus of theoretical studies
of the SSM \cite{Lee19} and other models with PSS and AFM phases
\cite{Zhao19,Sun21}. However, a recent density matrix renormalization
group (DMRG) study detected a gapless QSL state intervening between
the PSS and AFM phases of the SSM \cite{Yang22} within a narrow range
of couplings, approximately $g \in (0.79,0.82)$. Subsequently, an
intervening phase with similar boundaries was also indicated by a
functional renormalization-group calculation \cite{Keles22}. If these
results are correct, they open the interesting possibility of a QSL
phase also between the PSS and AFM phases in SCBO, somewhere in the
pressure range $2.6$ to $3.2$ GPa, where experiments so far
\cite{Guo20,Jimenez20} have not detected any conventional phase
transitions or long-range order.  This prospect of realizing a gapless
QSL is especially important considering that SCBO can be synthesized
with very low concentration of impurities, thus, it is free of the
complicating disorder effects mentioned above.

The aim of the present work is to further corroborate the QSL phase
argued in Ref.~\onlinecite{Yang22}, where excited-state gaps computed
with the DMRG method were analyzed. Gap crossings associated with
quantum phase transitions were identified, similar to the previously
studied $J_1$-$J_2$ square-lattice Heisenberg model \cite{Wang18}
(where several other works also agree on the existence of a QSL in
roughly the same coupling range \cite{Gong14,Morita15,Ferrari20,Nomura20,Schackleton21}).
Crossing points flowing with increasing system size to two different
points were found, $g_{\rm c1} \approx 0.79$ and $g_{\rm c2} \approx 0.82$, and
these were associated with transitions out of the PSS phase and
into the AFM phase, respectively. The gaps and correlation functions in the
window $[g_{\rm c1},g_{\rm c2}]$ supported a gapless QSL phase
between the PSS and AFM phases.

Here we develop an improved level-spectroscopy
method, using combinations of excitation energies beyond the gaps
with respect to the ground state. By judicial choices of quantum
numbers and identification of composite and elementary excitations
on fully periodic lattices, spectral gap combinations can be defined
whose crossing points exhibit only very weak dependence on the lattice size. 
Even with the small clusters accessible with exact diagonalization, we can 
confirm crossing points in excellent agreement with those extrapolated from 
the conventional gap crossings in much larger systems with cylindrical boundary 
conditions \cite{Yang22}.  To further demonstrate the improved gap crossing
method, we also consider a spin chain with long-range interactions, illustrated
in Fig.~\ref{hh}(b), which has 
a similar ground-state phase structure as a function of an exponent controlling 
the long-range couplings.

We also study the relevant order parameters of both models. The results 
further demonstrate the utility of the level crossing method to detect quantum
phase transitions when the system sizes are too small to reliable extrapolate the 
order parameters to the thermodynamic limit.

{\bf Exact diagonalization and level crossings}---Exact
diagonalization of the Hamiltonian is the most versatile numerical
method for quantum lattice models, however strongly limited to small
lattice sizes owing to the prohibitive exponential growth of the
Hilbert space. Proper selection of cluster sizes and shapes, and
thorough examination of their lattice symmetries (conserved quantum
numbers for block-diagonalization), are the two most
important steps for fully utilizing the power of the method
\cite{Laflorencie04,Noack05,Weisse08,Lauchli11,Sandvik10ed}.  

The quantum numbers are also important for understanding and exploiting
excitations, which are useful not only in their own right but also for
detecting phase transitions of the ground state.  The underlying
assumption of the level spectroscopic method that we will use here is
that a change in the ground state at a quantum phase transition is
also accompanied by a change in the elementary excitations, which can
be reflected in a re-arrangement of energy levels with different
quantum numbers. If that is the case, there will be real level
crossings of excited states even when ground state transition
takes place through an avoided level crossing (i.e., with the quantum
numbers of the ground state on a finite cluster not changing versus
the control parameter).

The level crossing method is very well known in the context of 1D
models, especially the frustrated $J_1$-$J_2$ Heisenberg chain where
this approach originated \cite{Nomura92,Eggert96}. The power of the
method in this case lies in the fact that the crossing point between
the lowest singlet and triplet excitations versus $J_2/J_1$ converges
very rapidly to the critical point with increasing chain length $N$, with
shifts proportional to $N^{-2}$.  Subleading corrections are small, and the
transition point can be obtained to precision $10^{-6}$
\cite{Eggert96} or even better \cite{Sandvik10ed} even with chain
lengths only up to $N=32$, easily accessible with exact
diagonalization. In other cases, e.g., the chain with long-range
interactions that we will also consider here, the subleading
corrections are more substantial but still reliable results can be
obtained with relatively small chains \cite{Sandvik10}.

More recently, the level-crossing approach has also been applied to
2D systems, in combination with a variety of methods for computing the
relevant excited states, e.g., quantum Monte Carlo \cite{Suwa16}, DMRG
\cite{Wang18}, and sophisticated variational wave functions
\cite{Ferrari20,Nomura20}.  In the previous application to the SSM
\cite{Yang22}, the DMRG method was used to generate excited states in
several symmetry sectors on cylindrical lattices (i.e., with open
boundaries in one lattice direction and periodic boundaries in the
other direction). With fully periodic lattices, results converged to
the degree necessary for reliable level-crossing studies are difficult
to obtain with the DMRG method for system sizes much beyond those
for which exact diagonalization (with, e.g., the Lanczos method) can
be used. Periodic lattices are preferable, because of their
higher symmetry, thus allowing access to additional quantum numbers
beyond those used with the DMRG method. Here we will show that even very
small periodic lattices already contain the spectral information
pertaining to the ground state transitions of the SSM, but suitable
gaps and combinations of gaps have to be identified.

The SSM Hamiltonian is
\begin{equation}
\label{ssm}
H=J\sum_{\langle ij\rangle}\mathbf{S}_i\cdot \mathbf{S}_{j} + J'\sum_{\langle ij\rangle'}\mathbf{S}_i\cdot \mathbf{S}_{j},
\end{equation}
where $\mathbf{S}_{i}$ are $S=1/2$ operators, $\langle ij\rangle$ in the $J$ sum represents
all nearest-neighbor site pairs on a 2D square lattice, and $\langle ij\rangle'$ in the $J'$ sum
indicates next-nearest neighbor sites belonging to one of the diagonal bonds in a given plaquette but
only in every second plaquette, as illustrated in Fig.~\ref{hh}(a). As is customary, we will refer
to the plaquettes with and without $J'$ interaction as filled and empty, respectively.

\begin{figure}
\includegraphics[width=80mm]{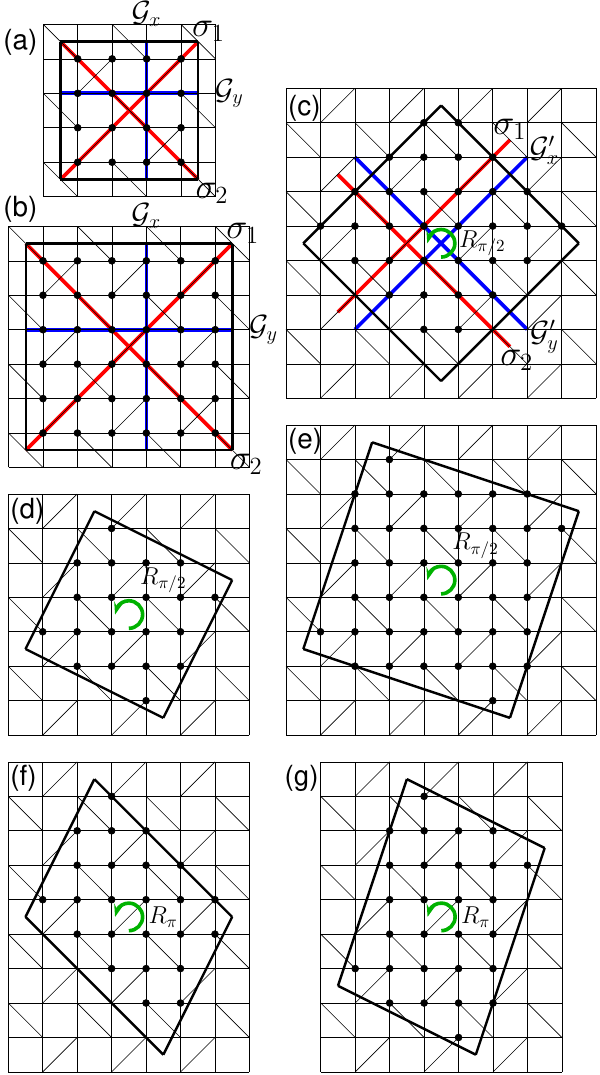}
\vskip-1mm
\caption{The seven clusters studied in this work. For each system size $N$, a cut-out from the infinite lattice is indicated
  and periodic boundary conditions are applied to these finite clusters. The clusters are arranged according to their different 
  symmetries and the sizes $N$ are (a) $16$, (b) $36$, (c) $32$, (d) $20$, (e) $40$, (f) $24$, and (g) $28$,
  with the sites included in each cluster marked by the black circles. The lattice symmetries are
  illustrated as follows: Gliding reflection operators $\mathcal{G}_x$ and $\mathcal{G}_y$ in (a) and (b), defined in
  Eq.~(\ref{gdef}), involve reflection with respect to the blue lines; analogous operations $\mathcal{G}'_x$ and $\mathcal{G}'_y$
  are defined for the cluster in (c). Mirror reflections $\sigma_1$ and $\sigma_1$ are defined with respect to the red lines in (a)-(c).
  Rotation $R_\phi$ by an angle $\phi$ is defined with respect to the center of an empty or filled plaquette as indicated by the
  green semi-circles in (c)-(g).} 
\label{ll}
\end{figure}

We use the Lanczos method for periodic clusters with $N=16,20,24,28,32,36$; see Fig.~\ref{ll}.
The $N=40$ system is beyond the reach of the Lanczos method within our
computational resources, but some of its low-energy states can be completely
converged by the implementation of the DMRG method described in Ref.~\cite{Yang22}.
For larger fully periodic clusters, convergence of excited states to the degree we demand here also
becomes too challenging for DMRG (in contrast to the much larger cylindrical lattices
studied previously \cite{Yang22,Lee19}).

The specific Heisenberg chain with long-range interactions that we also study as a benchmark
case is defined by
\begin{equation}
\label{lrchain}
H=\sum_{i=1}^N\sum_{r=1}^{N/2}J_r \mathbf{S}_i\cdot \mathbf{S}_{i+r},
\end{equation}
where the distance dependent couplings are given by \cite{Sandvik10}
\begin{equation}
\label{lrchain2}
J_2=g,\quad J_{r\neq 2}=\frac{(-1)^{r-1}}{r^{\alpha}}\left (1+\sum_{r^{\prime}=3}^{N/2}\frac{1}{r^{\prime\alpha}}\right)^{-1},
\end{equation}
with adjustable parameters $\alpha$ and $g$ and the normalization of
$J_{r\not=2}$ chosen such that the sum of all nonfrustrated
($r\not=2$) interactions $|J_r|$ equals $1$. This model has been
studied in previous works using the conventional level-crossing
approach with energies computed with the Lanczos method for $N$
up to $32$~\cite{Sandvik10ed} as well as with DMRG (in this case with
fully periodic boundary conditions) with $N$ up to $48$~\cite{Wang18}.

The existence of a gapless QSL in this 1D model is not controversial, as
even the Heisenberg chain with only nearest-neighbor interactions has a
disordered ground state with algebraically decaying correlations. With
the long-range un-frustrated interactions, long-range AFM order stabilizes
when $\alpha$ is below a critical value close to $2$, with the exact value
depending on short-distance details of $H$ \cite{Laflorencie05}. The third
phase in this case is the same frustration-driven two-fold degenerate 
dimerized phase as in the $J_1$-$J_2$ chain. The QSL can be expected on
general grounds for some range of the model parameters to be located
between the AFM and dimer phases, and this was confirmed in
Refs.~\cite{Sandvik10,Wang18}. Here we will show that the improved
level crossing method that we developed for the SSM produces better results
for the chain Hamiltonian Eq.~(\ref{lrchain}) as well. The behavior of various
gap crossing points, as the system transitions from dimerized to QSL and then to
AFM, are very similar to those observed in the SSM.

{\bf Symmetries of the SSM}---The lattice symmetries exploited here
are illustrated in Fig.~\ref{ll} for all the SSM clusters used in our
study. These symmetries are used to block diagonalize the Hamiltonian
along with the conserved magnetization $S^z$ and the spin-inversion symmetry 
$Z$ (the latter only for $S^z=0$ states). We do not use the total spin $S$ for 
block diagonalization, because of the complicated basis vectors in this case, 
but we compute $S$ of the eigenstates after the diagonalization procedure.

We first discuss the point-group symmetries of the standard $4\times 4$ ($N=16$)
and $6\times 6$ ($N=36$) clusters; see Figs.~\ref{ll}(a,b). These clusters have
translational symmetry in the ${x}$ and ${y}$ lattice directions, which
we define using the operators
\begin{equation}
\mathcal{T}_x=T^2_x,~~~~ \mathcal{T}_y=T^2_y,
\label{tdef}
\end{equation}
where $T_x$ and $T_y$ denote the operations of
translating by one lattice spacing in the respective directions. Periodic boundaries
for an $L \times L$ cluster with even $L$ imply the conditions 
$\mathcal{T}^{L/2}_x=\mathcal{T}^{L/2}_y=1$.

We use the gliding reflection symmetries defined by
\begin{equation}
\mathcal{G}_x=T_yP_x,~~~~ \mathcal{G}_y=T_xP_y,
\label{gdef}
\end{equation}
where $P_x$ and $ P_y$ are mirror (reflection) operations with respect to vertical
and horizontal lines passing through lattice sites. We also use diagonal
mirror reflections $\sigma_1$ and $\sigma_2$, defined with respect to lines drawn
through intra-dimer ($J'$) bonds. The $L \times L$ clusters are also invariant
under the composite rotation defined as
\begin{equation}
\mathcal{R}=T_xT_yR_{\pi/2}=\mathcal{G}_x\sigma_1,
\label{rdef}
\end{equation}
where $R_{\pi/2}$ is the 90$^{\circ}$ rotation operation, but this  composite symmetry does not further
reduce the size of the Hamiltonian blocks after the other symmetries have been used. We nevertheless compute
the eigenvalue of $\mathcal{R}$ using that of $\mathcal{G}_x$ and $\sigma_1$.

The $N=32$ cluster is contained in a square that is $45^{\circ}$ rotated with respect to the
lattice axes; see Fig.~\ref{ll}(c). Defining $T^{\prime}_x$ and $T^{\prime}_y$ as translations along
the diagonal directions by one step, the cluster is invariant under the following operations:
$\mathcal{T}^{\prime}_x=T^{\prime 2}_x$,
$\mathcal{T}^{\prime}_y=T^{\prime 2}_y$, $\mathcal{G}^{\prime}_x=T^{\prime}_yP^{\prime}_x$,
$\mathcal{G}^{\prime}_y=T^{\prime}_xP^{\prime}_y$, $\sigma_1$,
$\sigma_2$, and $R_{\pi/2}$. Here we have defined $P^{\prime}_x$ and
$P^{\prime}_y$ as mirror operations with respect to diagonal lines passing only through empty
plaquettes. Imposing periodic boundary conditions corresponds to 
$\mathcal{T}^{\prime 2}_x=\mathcal{T}^{\prime 2}_y=1$. For this cluster, the rotation 
symmetry $R_{\pi/2}$ is also useful for block diagonalization.

The $N=20$ and $N=40$ clusters, Fig.~\ref{ll}(d,e), are invariant under $\mathcal{T}_x$ and
$\mathcal{T}_y$, and because of the tilting the periodicity implies $\mathcal{T}_x^2\mathcal{T}_y=1$
for $N=20$ and $\mathcal{T}_x^3\mathcal{T}_y=1$ for $N=40$. We also use the $90^\circ$ rotation symmetry,
$R_{\pi/2}$, with respect to the center of an empty plaquette.

Finally, the $N=24$ and $N=28$ clusters, Fig.~\ref{ll}(f,g), are similar, being symmetric with respect
to a $180^\circ$ rotation $R_{\pi}$ about the the center of a filled plaquette. The translational constraints
are $\mathcal{T}_x\mathcal{T}_y^2=1$ and $\mathcal{T}_x^2\mathcal{T}_y=1$, respectively, for $N=24$ and
$N=28$.

\begin{table}
\begin{center}
 \caption{\label{tab_sec} Quantum numbers corresponding to the various point-group and spin symmetries for the investigated low-energy states of
 clusters with $N=16$, $N=32$, and $N=36$. All states have quantum number $+1$ (momentum zero) of the applicable translations    
 $\mathcal{T}_x, \mathcal{T}_y$ or $\mathcal{T}^{\prime}_x, \mathcal{T}^{\prime}_y$. The spin inversion symmetry $Z$ is used only when $S^z=0$.}
\vskip2mm
 \begin{tabular}{|r|r|r|r|r|r|r|r|r|}
\hline
& $\mathcal{G}_x$,$\mathcal{G}^{\prime}_x$ & $\mathcal{G}_y$,$\mathcal{G}^{\prime}_y$ & $\sigma_1$ & $\sigma_2$ & $\mathcal{R}$,$R_{\pi/2}$ & $S^z$ & $S$ & $Z$
\tabularnewline \hline
$S_1$ & 1~~~ & 1~~~ & 1~ & 1~ & 1~~~~ & 0~ & ~0~ & 1~
\tabularnewline \hline 
$S_2$ & -1~~~ & -1~~~ & -1~ & -1~ & 1~~~~ & 0~ & ~0~ & 1~
\tabularnewline \hline 
$T_1$ & -1~~~ & -1~~~ & 1~ & 1~ & -1~~~~ & 0~ & ~1~ & -1~
\tabularnewline \hline 
$T_2$ & 1~~~ & 1~~~ & -1~ & -1~ & -1~~~~ & 0~ & ~1~ & -1~
\tabularnewline \hline
$Q_1$ & 1~~~ & 1~~~ & 1~ & 1~ & 1~~~~ & 2~ & ~2~ & $/$~
\tabularnewline \hline 
\end{tabular}
\end{center}
\end{table}

\begin{table}
\begin{center}
 \caption{\label{tab_sec2} Quantum numbers of the investigated state with respect to the applicable rotations for $N=20$, $N=24$, and $N=28$
 clusters. All states have momentum zero.} 
\vskip2mm
 \begin{tabular}{|r|r|r|r|r|r|}
\hline
& $R_{\pi/2}$ ($N=20$) & $R_{\pi}$ ($N=24,28$) & $S^z$ & $S$ & $Z$
\tabularnewline \hline
$S_1$ & 1~~~~ & 1~~~~ & 0~ & ~0~ & 1~
\tabularnewline \hline 
$S_2$ &  1~~~~ & ~1~~~~ & ~0~ & ~0~ & 1~
\tabularnewline \hline 
$T_1$ & -1~~~~ & ~1~~~~ & ~0~ & ~1~ & -1~
\tabularnewline \hline 
$T_2$ & -1~~~~ & ~1~~~~ & ~0~ & ~1~ & -1~
\tabularnewline \hline
$Q_1$ & 1~~~~ & ~1~~~~ & ~2~ & ~2~ & $/$~
\tabularnewline \hline 
\end{tabular}
\end{center}
\end{table}

{\bf Characteristic SSM eigenstates}---Upon increasing $g$, the SSM undergoes
a first-order quantum phase transition between the unique DS state and the two-fold degenerate
PSS state by a true level crossing 
at $g\approx 0.685$ \cite{koga00,Corboz13}. We here focus solely on changes in the low-energy
level spectrum for $g \ge 0.7$, excluding the well understood DS phase and the trivial transition 
out of it. We target the quantum phase transition from the PSS ground state to the
putative QSL state at $g = g_{\rm c1} \approx 0.79$, followed by the transition from this state into the
AFM state at $g = g_{\rm c2} \approx 0.82$ \cite{Yang22}. Thus, we aim to understand how the low-energy
spectrum changes as a function of $g$, as in Ref.~\cite{Yang22} but with important differences
because of the cylindrical boundary conditions used previously and the fully periodic
clusters studied here.

The two-fold degenerate singlet ground state is an essential and useful feature of the PSS phase
of the SSM on the fully periodic clusters studied here. We label these states, whose degeneracy is lifted by
finite-size effects, as $S_1$ and $S_2$. The characteristic Anderson rotor tower of states \cite{Anderson59}
is a hallmark of AFM order, and we consider the first two of these multiplets; the triplet excitation
$T_1$ (which we compute in the $S^z=0$ sector) and the quintuplet $Q_1$ (for practical reasons computed 
in the $S^z=2$ sector). The intermediate QSL state of the SSM argued in Ref.~\cite{Yang22}
has not yet been fully characterized, and, thus, there are no rigorously known distinguishing spectral
features of it. However, the results of Ref.~\cite{Yang22} indicate that it should have gapless singlet
and triplet excitations. Thus, all three phases under consideration should have gaps that vanish as the
system size is increased, and we are interested in potential level crossings signaling the ground
state phase transitions.

In addition to the four low-energy states $S_1$, $S_2$, $T_1$, and $Q_1$, discussed above,
we also study a triplet $T_2$ that can be regarded as an excitation above $S_2$ with the
same relative quantum numbers as those of $T_1$ relative to $S_1$. All states studied here have
momentum zero, i.e., the phase factor generated when applying the translation operators $\mathcal{T}_x$
and $\mathcal{T}_y$ in Eq.~(\ref{tdef}) to these states is $+1$. The absolute and relative lattice quantum 
numbers of interest here are therefore only the even ($+1$) and odd ($-1$) phases associated with the point-group 
symmetry operations. The absolute quantum numbers of the $N=16,32$ and $N=36$
clusters are listed in Tab.~\ref{tab_sec}, and in Tab.~\ref{tab_sec2} the applicable quantum numbers are
similarly listed for $N=20$ and $N=24$, and $28$. For $N=40$, we have not been able to converge the target
state $T_2$ with DMRG, but for all other states the quantum numbers are the same as those for $N=20$.

\begin{figure}
\includegraphics[width=55mm]{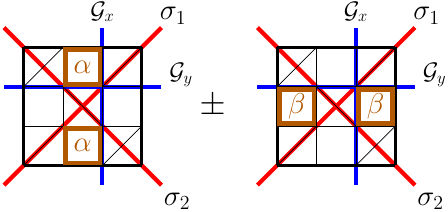}
\vskip-1mm
\caption{Cartoon picture of the $\pm$ superpositions of $\alpha$ type (bold
  squares in the left configuration) and $\beta$ type 
  (right configuration) singlet plaquettes that form the two-fold degenerate  
  ground states (quasi-degenerate for finite $N$) $S_1$ ($+$) and $S_2$ ($-$)
  of the PSS phase. Some of the symmetry operations used to understand (as explained
  in the text) the quantum numbers of the low-energy excitations $T_1$, $T_2$, and $Q_1$
  are indicated with corresponding mirror lines.}
\label{gsdemo}
\end{figure}

The listed quantum numbers in Tabs.~\ref{tab_sec} and \ref{tab_sec2} can be understood with
the aid of a cartoon picture of the two lowest singlet states in the PSS phase, illustrated in
Fig.~\ref{gsdemo}. These quasi-degenerate ground states of a finite cluster, which do not break the
two-fold order-parameter symmetry, are even ($S_1$) and odd ($S_2$) superpositions of the two different 
plaquette tilings (with singlets on empty plaquettes, as is the case in the SSM \cite{Corboz13})
that we refer to as  $\alpha$ and $\beta$. In Fig.~\ref{gsdemo}, only two singlet plaquettes
on empty squares are highlighted for each case (i.e., those that fit within the
small $4\times 4$ cluster). Though the SSM Hamiltonian is not bipartite, below we will
also invoke the checkerboard sublattices A and B of the square-lattice sites.

First consider operation on the $S_1$ or $S_2$ state by either $\mathcal{G}_x,\mathcal{G}_y,\sigma_1$, or $\sigma_2$
on the clusters in Fig.~\ref{ll}. All these operations effectively exchanges the $\alpha$ and $\beta$ sets of singlet plaquettes,
therefore generate a phase (quantum number) $+1$ and $-1$ when acting on the $S_1$ and $S_2$, respectively, thus explaining
the corresponding quantum numbers listed in Tab.~\ref{tab_sec}.

To understand the  quantum numbers of the triplet excitations, $T_1$ and $T_2$, first note that a plaquette
singlet can be regarded as a superposition of two parallel two-spin singlet bonds. Each singlet
bond connects the A and B sublattices, thus, are odd with respect to exchanging A$\leftrightarrow$B
of the two sublattices. For a system in which the total number of singlet bonds is even, i.e., for $N$ 
being an integer multiple of four (which is the case for all clusters studied here), the total product wave
function of these singlets is even under A$\leftrightarrow$B. If one singlet is excited to a triplet,
which is even under A$\leftrightarrow$B, such a state is anti-symmetric with respect to sublattice exchange. 
Note further that the operators $\mathcal{G}_x,\mathcal{G}_y$ involve A$\leftrightarrow$B site exchange 
while  $\sigma_1$ and $\sigma_2$ do not. Thus, the quantum number $-1$ of
$\mathcal{G}_x$ and $\mathcal{G}_y$ in the $T_1$ state arises from swapping A$\leftrightarrow$B
because there is an odd number of remaining singlets pairs. Similarly, the quantum number
$+1$ for $\sigma_1$ and $\sigma_2$ in $T_1$ follows because there is no sublattice swap. The same reasoning
applies to the state $T_2$, i.e., the triplet excitation of $S_2$; the relative sign difference in
the gliding and mirror quantum numbers with respect to $T_1$ (Tab.~\ref{tab_sec}) arises from the
odd superposition of the two sets $\alpha,\beta$ of plaquette tilings in $S_2$.

\begin{figure}
\includegraphics[width=\columnwidth]{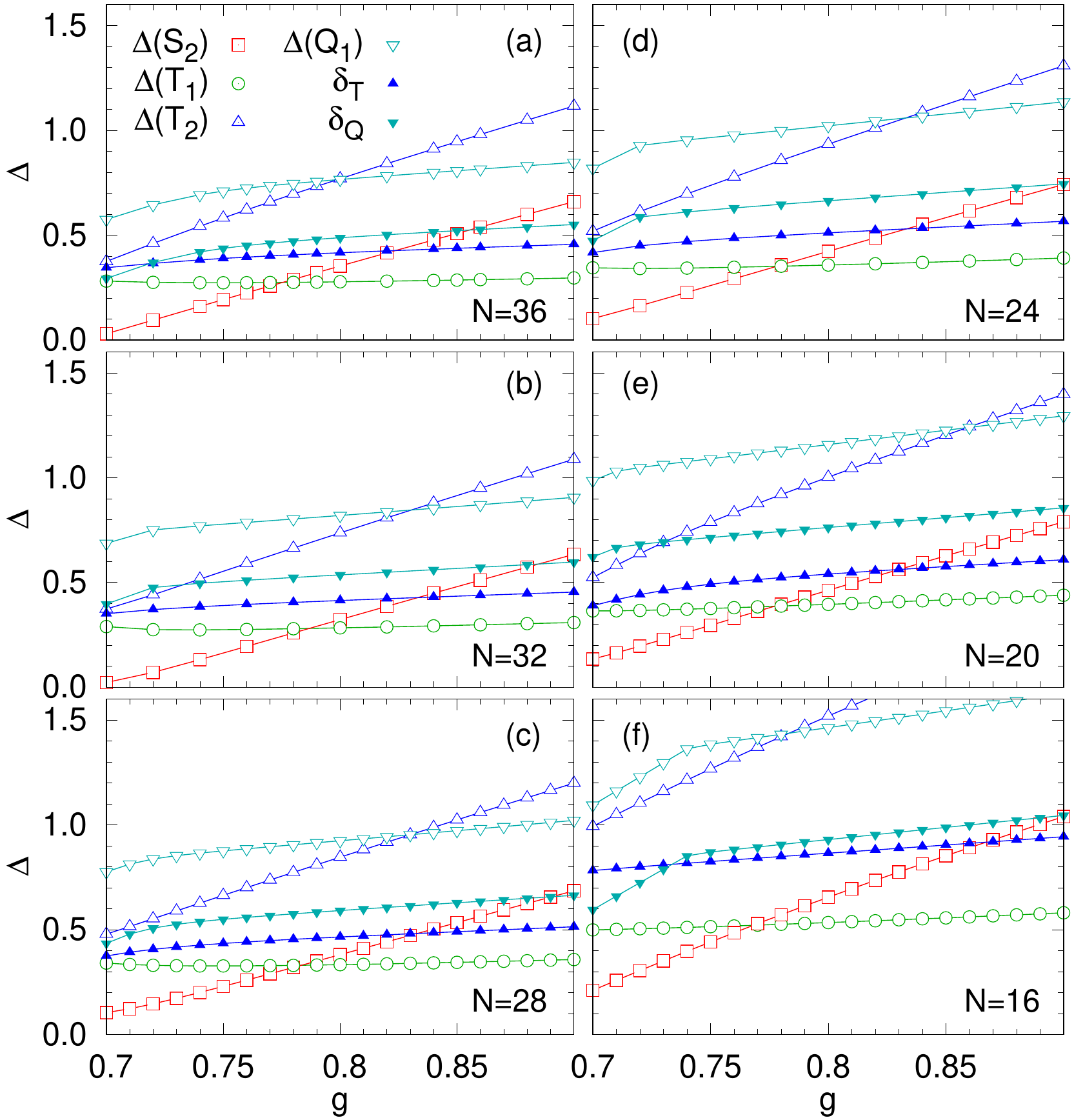}
\vskip-1mm
\caption{Energy gaps of the SSM vs the coupling for cluster sizes (a) $N=36$ (b) $32$,
  (c) $28$, (d) $24$, (e) $20$, (f) $16$. Conventional gaps defined relative to the ground state energy
  $E(S_1)$ are shown as follows: $\Delta(S_2)$ (open red squares), $\Delta(T_1)$ (open green circles),
  $\Delta(T_2)$ (open blue up triangles), $\Delta(Q_1)$ (open indigo down triangles).
  Triplet and quintuplet gaps defined with respect to other exited states are shown as follows:
  $\delta_T \equiv E(T_2)-E(S_2)$ (filled blue up triangles); $\delta_Q = E(Q_1)-E(T_1)$
  (filled indigo down triangles). The kinks in the $\Delta(Q_1)$ and $\delta_Q$ data between
  $g=0.7$ and $0.75$ are related to avoided level crossings close to the DS--PSS transition.}
\label{gapall}
\end{figure}

The state $Q_1$ can be thought of as the result of exciting two singlet dimers of $S_1$ into triplets, and by
applying symmetry operations as above, all reflection quantum numbers remain the same in $Q_1$ as in $S_1$
because of the even number of triplets.

The quantum numbers of the rotation operators, $\mathcal{R}$, $R_{\pi}$, or $R_{\pi/2}$, depending on the cluster,
can likewise be understood in light of Fig.~\ref{gsdemo} and how the symmetry operations correspond or
not to sublattice and plaquette swaps. As an example, for the $N=32$ cluster the rotation operator $R_{\pi/2}$ 
swaps the A and B sublattices but not the $\alpha$ and $\beta$ singlet plaquettes. Therefore, for
the states $S_1,S_2$ and $Q_1$, which contain an even number of singlet bonds, the quantum number
is $+1$, while for $T_1$ and $T_2$, which contain an odd number of singlets, the rotation quantum
number is $-1$.

The above arguments apply to all clusters in Fig.~\ref{ll} with their respective applicable symmetry operations.
We have explained the quantum numbers by examining a simple picture of the singlets in the PSS phase, and when moving 
to other phases the energy levels for the finite systems evolve continuously. The states $\{S_1,S_2,T_1,T_2,Q_1\}$ 
are still defined according to their quantum numbers listed in Tabs.~\ref{tab_sec} and \ref{tab_sec2} and are always
those that evolve from the two lowest singlets, two lowest triplets, and lowest quintuplet in the PSS state.
The state $S_1$ remains the ground state for all values of $g$ considered, and $S_2$, $T_1$, and $Q_1$ 
also remain the lowest states with their respective total spin. However, $T_2$ is not always the second lowest
triplet in the QSL and AFM phases, though it is the first triplet with its full set of quantum numbers.

{\bf Numerical SSM results}---We define the gaps $\Delta(S_2)$, $\Delta(T_1)$, $\Delta(T_2)$, and $\Delta(Q_1)$
relative to the ground state energy $E(S_1)$ and graph these versus $g$ in Fig.~\ref{gapall} for the
clusters of size up to $N=36$. As explained above, our goal is to identify level (gap) crossings with the
PSS--QSL and the QSL--AFM ground state transitions.

In Ref.~\cite{Yang22}, the extrapolated (with leading $1/N$ corrections) crossing point $g_{\rm c1} = 0.788 \pm 0.002$ between the lowest
singlet and triplet excitation was identified as the PSS--QSL transition. Unlike the periodic clusters considered here, the cylindrical
lattices studied in Ref.~\cite{Yang22} break the asymptotic two-fold degeneracy of the PSS state because the boundaries favor one of the two singlet
patterns. Thus, the first excited singlet was different from the quasi-degenerate ground state $S_2$ used here, and the level crossing studied
previously is not a directly analogy to the singlet-triplet crossing accompanying
the dimerization transition in the frustrated Heisenberg chain \cite{Nomura92,Eggert96}
(where the symmetry is not broken in periodic systems). An important aspect of the present work is that the crossing between the $S_2$ and $T_1$ levels is
similar to the well understood 1D case, and a confirmation of the same asymptotic crossing point as in Ref.~\cite{Yang22} will represent additional
independent evidence for the correct identification of the quantum phase transition.

\begin{figure}
\includegraphics[width=65mm]{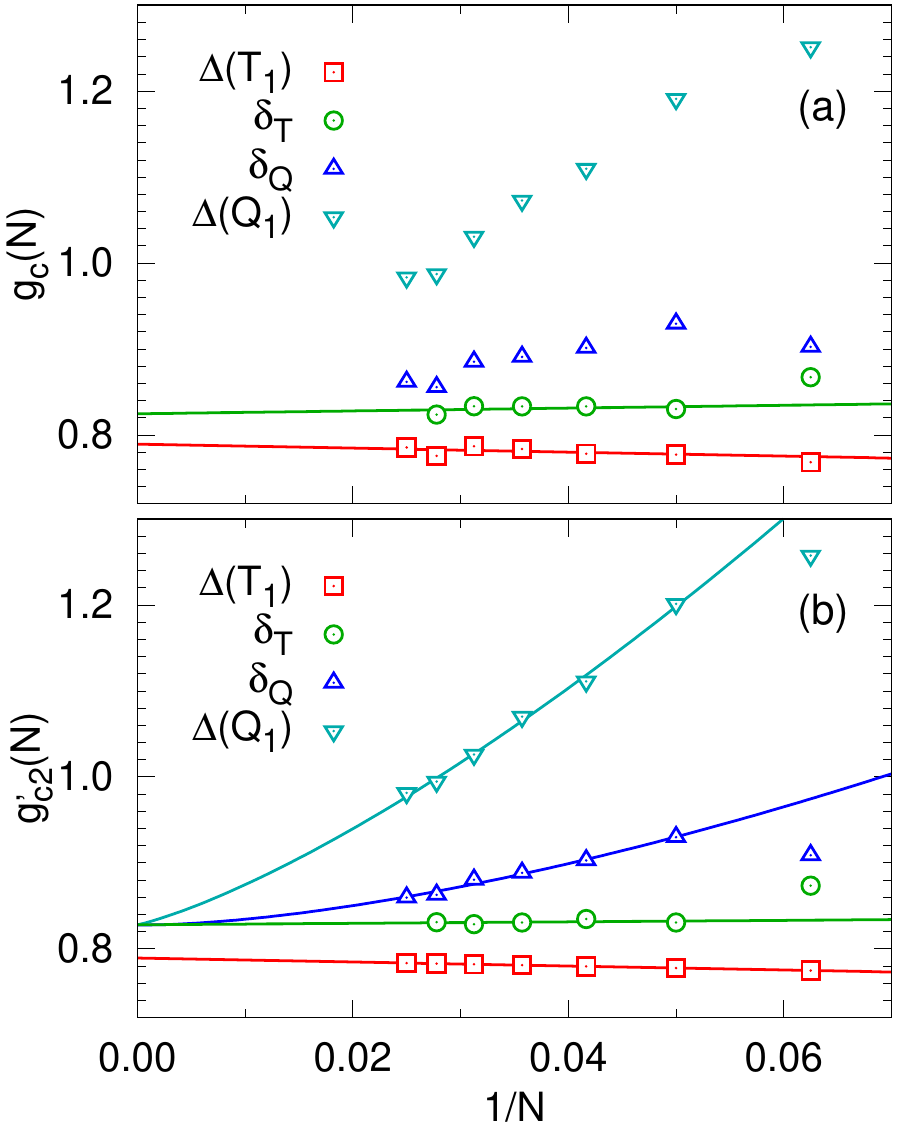}
\vskip-1mm
\caption{(a) SSM finite-size level crossing points obtained from the gaps
  $\Delta(T_1)$, $\delta_T$, $\delta_Q$, and $\Delta(Q_1)$, each crossing the
  singlet gap $\Delta(S_2)$. The points are graphed vs the inverse system size
  according to the empirical linear scaling in $1/N$ \cite{Wang18,Yang22}.
  The underlying data are from Lanczos calculations such as those in Fig.~\ref{gapall},
  except for the largest cluster, $N=40$, for which the DMRG method was used.
  The two straight lines are fits to the $\Delta(T_1)$ (red solid line)
  and $\delta_T$ (green solid line) points for $N\geq 20$ and extrapolate
  to $g_{c1}=0.789$ and $g_{c2}=0.824$, respectively. (b) Adjusted crossing points,
  $g_{c2}^{\prime}$, Eq.~(\ref{shiftedg}), for which all points for given $N$ are shifted vertically
  by an equal amount so that the $\Delta(T_1)$ points (red squares) fall exactly on the red 
  fitted line from (a). A linear fit (green line) in $1/N$ is   shown for the $\delta_T$
  crossing points and extrapolates to $g_{c2} = 0.826$. The form $g_{c2}^{\prime}(N)=g_{c2}+a/N+b/N^{3/2}$
  was fitted to the other two data sets ($N\geq 20$) with $g_{\rm c2}$ constrained to
  the same value as above.}
\label{criticalpoints}
\end{figure}

In Fig.~\ref{gapall}, the crossing of the $\Delta(S_2)$ and $\Delta(T_1)$ gaps indeed are also close to the previous $g_{\rm c1}$
value for all clusters. Interpolated crossing $g$ values are graphed versus $1/N$ in Fig.~\ref{criticalpoints}(a) (red squares), where we include also the $N=40$
result obtained with the DMRG method. Here the overall size dependence is much weaker than in the cylindrical lattices \cite{Yang22}, though there is some
un-smoothness as a consequence of the different cluster shapes. A line fit to all but the $N=16$ point gives $g_{\rm c1} = 0.789 \pm 0.004$ (where the estimated
 error, here and in other extrapolations reported below, was obtained from additional fits to all data sets excluding one of the points), in remarkable
agreement with the value cited above from the much larger cylindrical lattices (up to $N=24\times 12$ spins). The weak size dependence of the crossing points
and the consistency of the two calculations illustrate the advantage of periodic boundary conditions and also confirm the quantum-critical point with
a different level crossing.

The extrapolated crossing point between the lowest singlet and quintuplet excitations, $g_{\rm c2} = 0.820 \pm 0.002$, was identified as the
QSL--AFM transition \cite{Yang22}. This crossing point had a much larger size dependence on the cylindrical lattices than the singlet-triplet
crossing. The larger size dependence is also seen with our small periodic clusters, where the crossing points between
$\Delta(S_2)$ and $\Delta(Q_1)$ are outside the range of Fig.~\ref{gapall}. The crossing values, graphed in Fig.~\ref{criticalpoints}(a) 
(indigo down triangles), are consistent with the value of $g_{\rm c2}$ cited above but are too scattered for a meaningful extrapolation. 

Physically, the singlet-quintuplet crossing is motivated by the Anderson tower of rotor states in the AFM phase. The $S=0$ ground state 
$S_1$ is the lowest of these states, whose gaps with respect to $E(S_1)$ scale as $S(S+1)/N$ for $S>0$ \cite{Anderson59}.
Other singlets, including $S_2$, have
energies above these rotor states (for any $S>0$ and sufficiently large $N$). The triplet $T_1$, which becomes the $S=1$ rotor state in the
AFM phase, already crosses from above to below $S_2$ at the PSS-QSL transition point $g_{\rm c1}$, as discussed above. There is no necessary
reason why $Q_1$ should fall below $S_2$ in the QSL phase, e.g., in a scenario of a deconfined phase the quintuplet should contain 
four excited spinons, while $S_2$ and $T_1$ should be two-spinon excitations. However, being the $S=2$ rotor state in the AFM 
phase, $Q_1$ has to be below  $S_2$ there. Thus, the $g$ value of the crossing between $\Delta(Q_1)$ and $\Delta(S_2)$ in the limit of infinite 
system size should coincide with the formation of AFM long-range order. The fact that the extrapolated crossing point $g_{\rm c2}$ indeed is 
larger than $g_{\rm c1}$ (in Ref.~\cite{Yang22} and further below) supports an extended QSL phase instead of a direct transition 
point between the PSS and AFM phases. 

Here our aim is to identify other gap crossings associated with the QSL--AFM transitions, in particular with 
the hope of reducing the size dependence and allowing reliable extrapolation of $g_{\rm c2}$ even with small clusters. We note that the
lower transition point $g_{c1}$, as obtained in Ref.~\cite{Yang22} and confirmed here, should not be controversial as it is close
to other estimates of the end of the PSS phase \cite{Corboz13,Lee19}---in particular, in Ref.~\cite{Lee19} the size dependence of
the point marking the upper PSS bound is consistent with our $g_{\rm c1}$ value.

To construct better $g_{\rm c2}$ estimators, we first observe that the second triplet gap $\Delta(T_2)$ in Fig.~\ref{gapall}
closely follows the singlet gap $\Delta(S_2)$, reflecting the fact that $T_2$ can be regarded
as a triplet excitation of $S_2$, in correspondence to the role of the
first triplet $T_1$ with respect to the ground state $S_1$. Given that $S_1$ and
$S_2$ are quasi-degenerate ground states in the PSS phase, the difference
\begin{equation}
\delta_T \equiv E(T_2)-E(S_2) \equiv \Delta(T_2)-\Delta(E_2)
\label{deltatdef}
\end{equation}
will also converge with increasing system size to the non-zero gap in this phase, and $\delta_T$ must
then be above the singlet splitting $\Delta(S_2)$ for sufficiently large $N$ (as is seen clearly in Fig.~\ref{gapall} 
for all clusters). As already discussed above, in the AFM phase $S_2$ must be above the low-lying Anderson $S>0$ 
rotor states. However, given that $S_2$ remains the lowest singlet excitation also in the AFM phase, it must also 
host long-range order and its own associated Anderson rotor tower. As $T_1$ is the lowest rotor excitation of
$S_1$, the composite excitation $T_2$ is the lowest rotor state excited from $S_2$. Thus, in the AFM phase 
$\delta_T \propto 1/N$ and $\delta_T < \Delta(S_2)$, which is also seen for larger $g$ values in Fig.~\ref{gapall}.

In the putative gapless QSL phase, we expect $S_2$ to still be the lowest excited singlet (which is
also found numerically) and $\Delta(S_2)$ should vanish with increasing $N$. Likewise, $\Delta(T_1)$
should vanish as $N \to \infty$. Both the singlet and triplet gaps were found to scale as $N^{-1/2}$
on cylinders in Ref.~\cite{Yang22}. We also expect such scaling of the gap of $T_2$ relative to $S_2$,
i.e., $\delta_T \propto N^{-1/2} $. If $\delta_T$ remains larger than $\Delta(T_1)$ and $\Delta(S_2)$ also inside the QSL
phase (as in the PSS phase), then the crossing point of $\Delta(S_2)$ and $\delta_T$ will
signal the QSL--AFM transition. While we have no formal proof of this behavior, on general grounds 
one can expect a composite excitation, such as $T_2$ excited from $S_2$, to be energetically
more costly than its analogous elementary excitation, here $T_1$ obtained from the ground state $S_1$.

These expectations are indeed borne out by the numerical crossing points between $\Delta(S_2)$
and $\delta_T$ in Fig.~\ref{criticalpoints}(a) (green circles), where we observe a surprisingly weak size dependence.
Fitting a line to the data graphed versus $1/N$ for $N \ge 20$, the extrapolated QSL--AFM transition point
is at $g = 0.824 \pm 0.008$, fully consistent with $g_{\rm c2} = 0.820 \pm 0.002$ obtained previously with
the larger cylindrical clusters. In this case, we do not have results for $N=40$, as the DMRG 
calculation  for $T_2$ also demands calculation of several other triplets between $T_1$ and $T_2$
(with different quantum numbers that  are not resolved in our DMRG implementation \cite{dmrgnote}).

For yet another gap crossing corresponding to the QSL--AFM transition, we can construct a quantity similar
to $\delta_T$, Eq.~(\ref{deltatdef}), based on the quintuplet state $Q_1$. In analogy with $T_2$ being an excitation 
of $S_2$, we can also regard $Q_1$ as a further excitation of $T_1$. Defining the corresponding relative gap as
\begin{equation}
\delta_Q\equiv E(Q_1)-E(T_1)\equiv\Delta(Q_1)-\Delta(T_1),
\label{deltaqdef}
\end{equation}
we can make the same kind of arguments as in the case of $Q_1$ and $S_2$ in the AFM phase, now
with $\Delta(Q_1)=3\Delta(T_1)$ asymptotically from the Anderson tower energies. Thus, asymptotically 
$\delta_Q \to 2\Delta(T_1)$ and we must have $\delta_Q < \Delta(S_2)$ in the AFM phase. Thus, we expect
that the QSL--AFM transition is associated with the asymptotic crossing of  $\delta_Q$ and $\Delta(S_2)$.
Such crossing points are indeed within the range of the graphs in Fig.~\ref{gapall}, 
and in Fig.~\ref{criticalpoints}(a) the size dependence of the crossing $g$ values based on $\delta_Q$ (blue 
up triangles) is significantly reduced below that of $\Delta(Q_1)$. Visually the points are consistent with an
asymptotic flow to $g_{\rm c2}$, though the behavior is not smooth enough for extrapolating reliably.

An interesting observation in Fig.~\ref{criticalpoints}(a) is that the conventional singlet-triplet
crossing points (red open squares) and the crossing of the singlet and $\delta_T$ (green open circles)
are highly correlated. Therefore, the distance between the points, i.e., asymptotically the size of the
QSL phase, has much less size dependence than the individual crossing $g$ values. Upon close inspection,
such correlations are also visible in the other $g_{\rm c2}$ estimates. In Fig.~\ref{criticalpoints}(b)
we exploit these correlations (which should arise from the cluster shape affecting all low-energy excitations
in a similar way) by plotting points that are shifted by equal amounts up or down for given $N$, so that the
$\Delta(T_1)$ points coincide exactly with the line fitted to those points in Fig.~\ref{criticalpoints}(a).
In other words, we cancel out the cluster-dependent correlation effects by focusing on the relative crossing points
but still taking into account the overall $g$ scale by adding the values corresponding to the line extrapolating
to $g_{\rm c1}$. This procedure defines adjusted crossing points
\begin{equation}
  \label{shiftedg}
  g_{c2}^{\prime}(N)=g_{c2}(N)-g_{c1}(N)+g^{\rm fit}_{c1}(N),
\end{equation}
where $g_{c1}(N)$ is the $\Delta(T_1)$ crossing point, $g^{\rm fit}_{c1}(N)$ the corresponding value from the line fit,
and $g_{c2}(N)$ is one of the other three crossing points. The so adjusted crossing points in Fig.~\ref{criticalpoints}(b)
have much smoother size dependence. A line fit to all but the $N=16$ data points in the case of $\delta_T$ gives $g_{\rm c2} = 0.826 \pm 0.003$.
This result is only slightly above the previous DMRG cylinder result; the calculations essentially agree within their estimated errors.

The adjusted $g_{c2}^{\prime}(N)$ data sets from the $\Delta(Q_1)$ and $\delta_Q$ crossing points in Fig.~\ref{criticalpoints}(b) are also
significantly smoothed compared to their original $g_{\rm c2}(N)$ values in Fig.~\ref{criticalpoints}(a), though the visibly large corrections
to the linear form, in combination with the small number of points, still make independent extrapolations with these data sets difficult.
However, by fixing the $N \to \infty$ value to that obtained from the $\delta_T$ linear fits and with $1/N^{3/2}$ corrections included
(corresponding to $1/L^3$ when expressed in cluster length $L$), the adjusted $\Delta(Q_1)$ and $\delta_Q$ data can both be fitted well,
thus lending support to all three crossing points flowing to the QSL--AFM transition.

The extrapolated $g_{\rm c2}$ value of course also depends on the line fit to the $\Delta(T_1)$ data and its extrapolated $g_{\rm c1}$ value.
However, the most important aspect of this analysis is that it leaves little doubt that there is a gap $g_{\rm c2}-g_{\rm c1} > 0$ between 
the two transition points, with the lines fitted to the $\Delta(T_1)$ and $\delta_T$ data only approaching each other marginally 
with increasing system size. It should be noted that the estimated size of the QSL phase, $g_{\rm c2}-g_{\rm c1}= 0.037 \pm 0.003$, 
is not dependent on the two individual line fits but can be obtained by a single line fit to the difference between the $\Delta(T_1)$ 
and $\delta_T$ points in Fig.~\ref{criticalpoints}(a).

{\bf Numerical results for the spin chain}---Next we use the 1D frustrated spin chain Hamiltonian, Eq.~(\ref{lrchain}), to further validate the conclusions
drawn for the 2D SSM. In Ref.~\cite{Sandvik10} the phase diagram was constructed based on Lanczos results for level crossings of gaps with respect to
the ground state as well as correlation functions. In the limit of large decay exponent $\alpha$ of the long-range interaction, the model reduces to
the well understood frustrated J$_1$-J$_2$ chain, where a quantum phase transition between the critical ground state (a QSL of the Luttinger-liquid
type) and a two-fold degenerate dimerized ground state (which we now refer to as a dimer-singlet-solid, DSS) takes place at $J_2/J_1 \approx 0.2411$
\cite{Nomura92,Eggert96,Sandvik10ed}. For $J_2=0$, the interactions are not frustrated, and a previous quantum Monte Carlo and field theory study of a
similar model detected a transition between the critical state and an AFM state upon lowering the long range exponent $\alpha$ \cite{Laflorencie05}. Long-range AFM order
is also intuitively expected when the interactions become strong enough at long distances so that the Mermin-Wagner theorem (which prohibits long-range
AFM order for short-range interacting 1D Heisenberg systems) is no longer valid and mean-field behavior sets in. Already based on these limiting behaviors, it is clear that the phase
diagram in the full parameter space $(J_2,h)$, where for convenience we have defined $h=\alpha^{-1}$, contains DSS, QSL, and AFM phases. These
phases can be traversed in said order, similar to the phases of the SSM versus $g$, by following appropriate paths in the parameter space.

Based on the previous results for the phase diagram \cite{Sandvik10}, we here study the phase transitions along a vertical cut with $J_2=0.3$ fixed and $h$ varied. 
The same line in the phase diagram was also already studied with the DMRG method on periodic chains up to length $N=48$ in Ref.~\cite{Wang18}, where level
crossings of the second singlet $S_2$ with the first triplet ($T_1$) and the first quintuplet ($Q_1$) were extrapolated to infinite size, resulting in
$h_{c1}=0.316$ and $h_{c2}=0.476$. We here wish to demonstrate that the finite-size effects are reduced when instead using the composite gaps defined
in Eqs.~(\ref{deltatdef}) and (\ref{deltaqdef}). We only present Lanczos calculations of chains of even length $N$ up to $N=32$.

\begin{figure}
\includegraphics[width=\columnwidth]{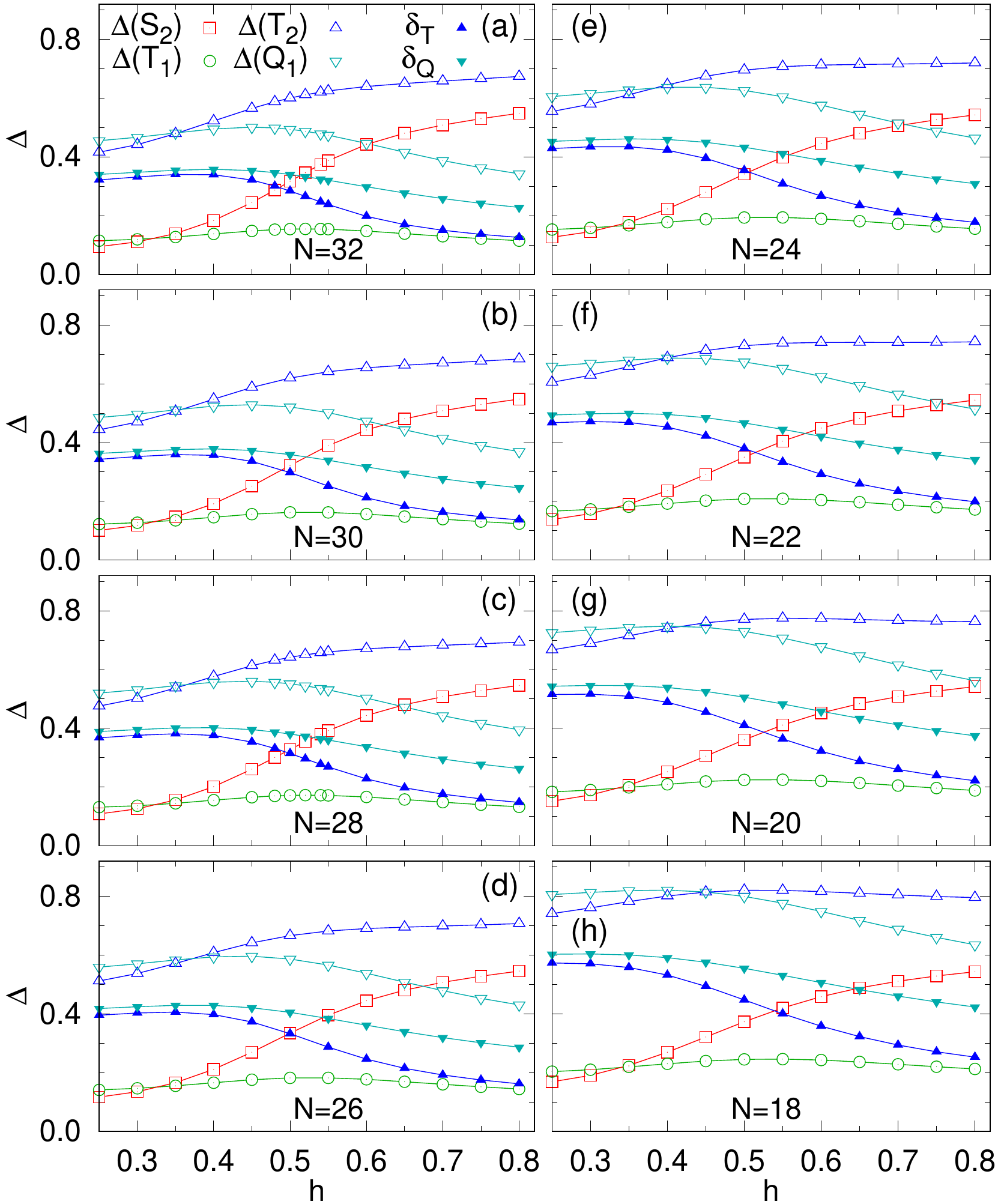}
\vskip-1mm
\caption{Energy gaps of the spin chain Hamiltonian in Eq.~(\ref{lrchain}) vs the inverse of the long-range exponent $h=\alpha^{-1}$: $\Delta(S_2)$ (open red squares),
  $\Delta(T_1)$ (open green circles), $\Delta(T_2)$ (open blue up triangles), and $\Delta(Q_1)$ (open indigo down triangles). The composite gaps
  are also shown; $\delta_T$ (filled blue up triangles) and $\delta_Q$ (filled indigo down triangles), as defined in Eqs.~(\ref{deltatdef}) and (\ref{deltaqdef}),
  respectively. The results in (a)-(h) were obtained with the Lanczos method for chains of length $N=32$ down to $N=18$ in steps of $2$.}
\label{gapchain}
\end{figure}

Figure \ref{gapchain} shows the four gaps relative to the ground state $S_1$ for chain sizes from $N=18$ to $N=32$.
The conventional gaps $\Delta(S_2)$, $\Delta(T_1)$, $\Delta(T_2)$, and $\Delta(Q_1)$ are shown along with
the two composite gaps $\delta_T$ and $\delta_Q$ defined in the same way as
in Eqs.~(\ref{deltatdef}) and (\ref{deltaqdef}). Here it should be noted that chains of length $N=4n$ (with integer $n$) have ground states with
momentum $k=0$, while $N=4n+2$ chains have $k=\pi$ (defined with translation by one lattice spacing).
The arguments that we made previously based on Fig.~\ref{gsdemo} regarding the quantum numbers
and physical interpretations of the low-energy states of the SSM apply also to the 1D model with its two-fold degenerate DSS in place of the PSS
of the SSM, including the dependence of the ground state momentum on the chain length (from even versus odd number of singlet bonds when $N=4n$
and $4n+2$, respectively). We therefore do not repeat the arguments for the level crossings flowing either to the DSS--QSL transition [$\Delta(T_1)$
crossing $\Delta(S_2)$] at $h=h_{\rm c1}$ or to the QSL--AFM transition [$\Delta(Q_1)$,  $\delta_T$, or $\delta_Q$ crossing $\Delta(S_2)$] at $h=h_{\rm c2}$.

\begin{figure}
\includegraphics[width=65mm]{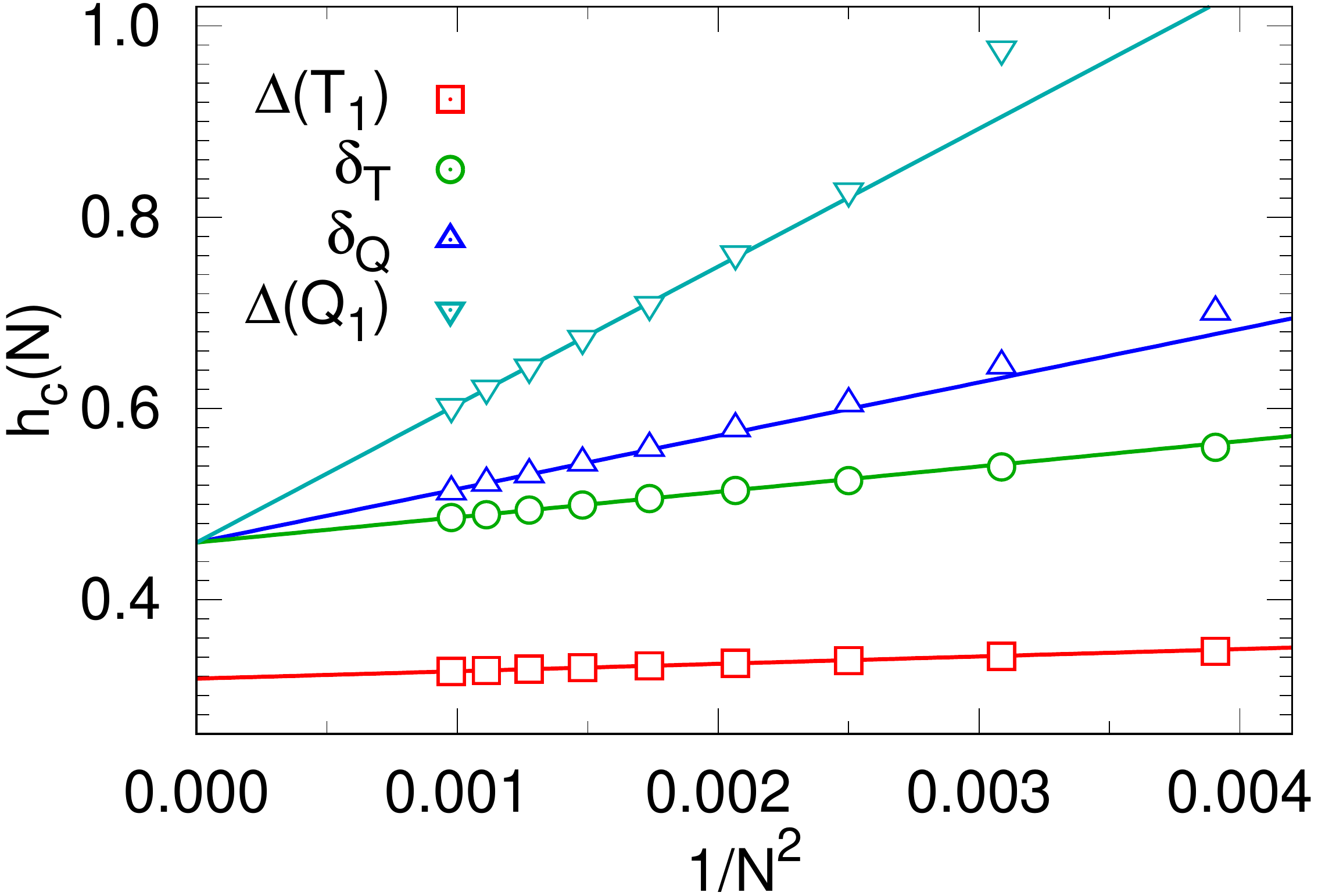}
\vskip-1mm
\caption{Finite-size crossing points extracted from the gap data in Fig.~\ref{gapchain}. The lines show fits in which a common
$N=\infty$ point was enforced in the case of $\delta_T$, $\delta_Q$, and $\Delta(Q_1)$. Data for small $N$ were excluded to obtain acceptable fits.}  
\label{criticalpoints2}
\end{figure}

In Fig.~\ref{criticalpoints2} we present the size dependence of the relevant crossing points, graphing them versus $1/N^2$ in which case we expect asymptotic
linear behavior. A clear window $h_{\rm c2}-h_{\rm c1} >0$ between the extrapolated crossing points is apparent here, corresponding to the known QSL phase
located between the DSS and AFM phases. Similar to the SSM, the $\Delta(T_1)$ crossing points only exhibit weak size dependence in their flow to the DSS-QSL
transition at $h_{\rm c1}$, here with smooth behavior as all system sizes correspond to the same shape of the lattice for all $N$. The extrapolated
transition point is $h_{\rm c1} = 0.3177 \pm 0.0002$, in good agreement with the previous results \cite{Wang18}, where the same level crossing was used
but with larger chains.

Also similar to the SSM, the weakest size dependence of the three estimates for the QSL--AFM transition point $h_{\rm c2}$ is achieved with $\delta_T$,
while $\Delta(Q_1)$ exhibits the largest variations with $N$. Thus, the use of a composite gap indeed also reduces the finite-size effects in this case,
In all cases, the data for the largest clusters can be fitted with lines versus $1/N^2$, with reasonable agreement between the different extrapolations and
in good agreement with the previous $h_{\rm c2}$ result. A constrained fit to all data sets (the lines shown in Fig.~\ref{criticalpoints2}) with a common
infinite-$N$ point gives $h_{\rm c2}=0.4600 \pm 0.0005$, which is slightly lower (about $3\%$) than the previous DMRG result and likely more reliable.

{\bf Order parameters}---We define the squared AFM (staggered) magnetization for both the SSM and the spin chain 
in the standard way as 
\begin{equation}
\label{msq}
  m_s^2=\frac{1}{N^2}\sum_{ij}\phi_{ij}\langle\mathbf{S}_i\cdot\mathbf{S}_j\rangle,
\end{equation}
where $\phi_{ij}=+1$ and $=1$ for sites $i,j$ in the same and different sublattices, respectively.

The squared dimer order parameter of the chain model is defined as
\begin{equation}
  \label{mdsq}
m_d^2=\frac{1}{N^2}\sum_{i,j}(-1)^{i-j}\langle D_iD_j\rangle,
\end{equation}
where $D_i=\mathbf{S}_{i}\cdot\mathbf{S}_{i+1}$. 

In the case of the SSM, singlets forming on the empty plaquettes can be detected with various operators. 
Here we define a plaquette operator solely with diagonal spin operators; 
$\Pi_i=\sigma^z_{i}\sigma^z_{i+\hat{x}}\sigma^z_{i+\hat{y}}\sigma^z_{i+\hat{x}+\hat{y}}$, where
$\sigma_i^z=2S^z_i$. Then
\begin{equation}
  \label{mpsqz}
m_{p}^2=\frac{4}{N^2}\sum_{i,j}\theta_{ij}\langle\Pi_i\Pi_j\rangle,
\end{equation}
where $i,j$ run only over the empty squares of the SSM lattice and $\theta_{ij}=+1$ and $-1$
for $i,j$ in the same row or different rows, respectively. 

\begin{figure}
\includegraphics[width=58mm]{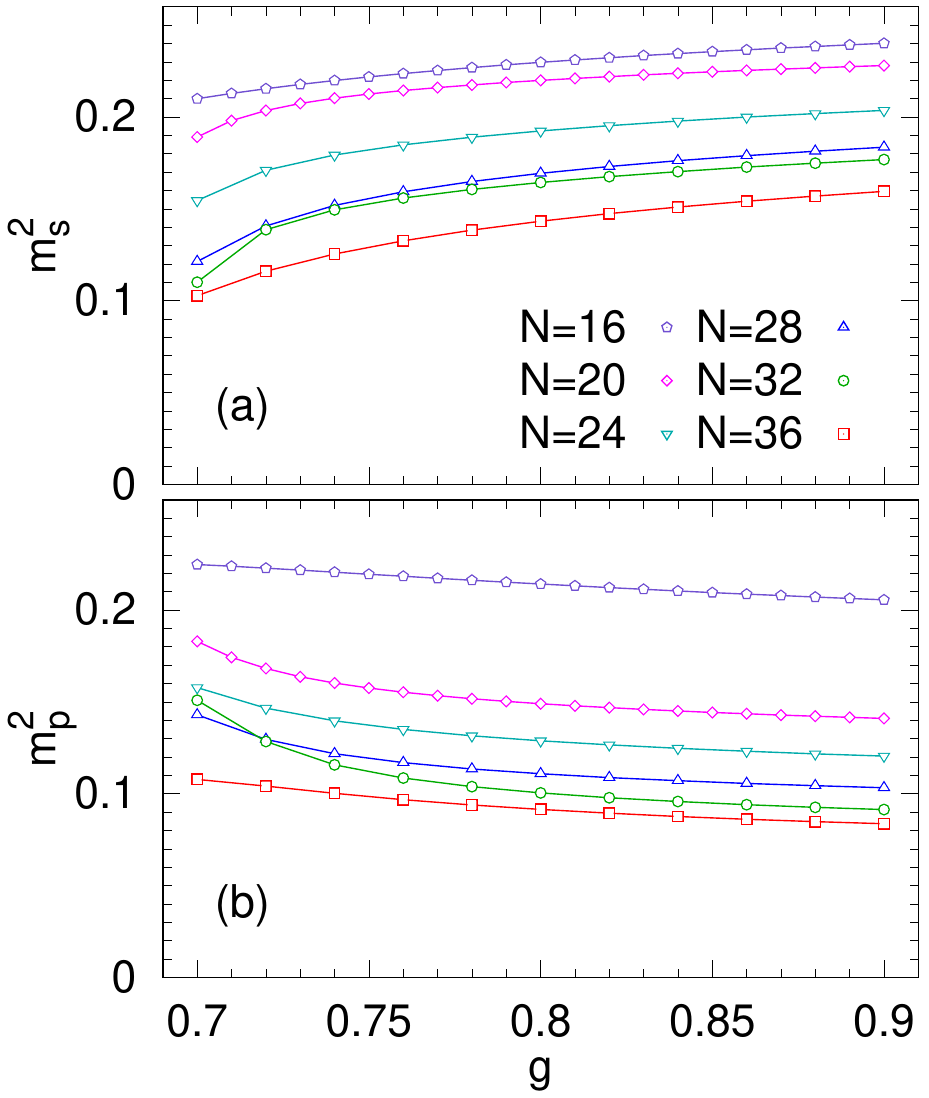}
\vskip-1mm
\caption{Squared order parameters of the SSM vs the coupling ratio for different system sizes. (a) Staggered magnetization, Eq.~(\ref{msq}),
(b) diagonal PSS order parameter, Eq.~(\ref{mpsqz}).}
\label{orderpara}
\end{figure}

Results for both order parameters of the SSM are shown in Fig.~\ref{orderpara}. While the AFM order
parameter increases with $g$ in Fig.~\ref{orderpara}(a) and the PSS order parameter $m^2_{p}$ correspondingly 
shows an overall reduction with $g$ in Fig.~\ref{orderpara}(b), the signals are clearly very weak. It is not possible 
to extrapolate these order
parameters to the thermodynamic limit, and we therefore do not show any such analysis here. In contrast, 
in the previous DMRG calculations \cite{Yang22} PSS and AFM order were clearly detected on the larger
cylindrical lattices in the relevant windows of $g$ values, and inside the QSL phase a power-law behavior 
of the AFM order parameter was observed. The system sizes accessible to the Lanczos method are simply too 
small for detecting the phase boundaries, or even to extrapolate the order parameters deep inside the PSS 
and AFM phases that certainly exist. We have also tried other definitions of the PSS order parameter, e.g., 
using cyclic permutation operators on the plaquettes instead of the diagonal operators $\Pi_i$ in Eq.~(\ref{mpsqz}), 
but the $g$ dependence is always weak, similar to the data in Fig.~\ref{orderpara}(b).

\begin{figure}
\includegraphics[width=60mm]{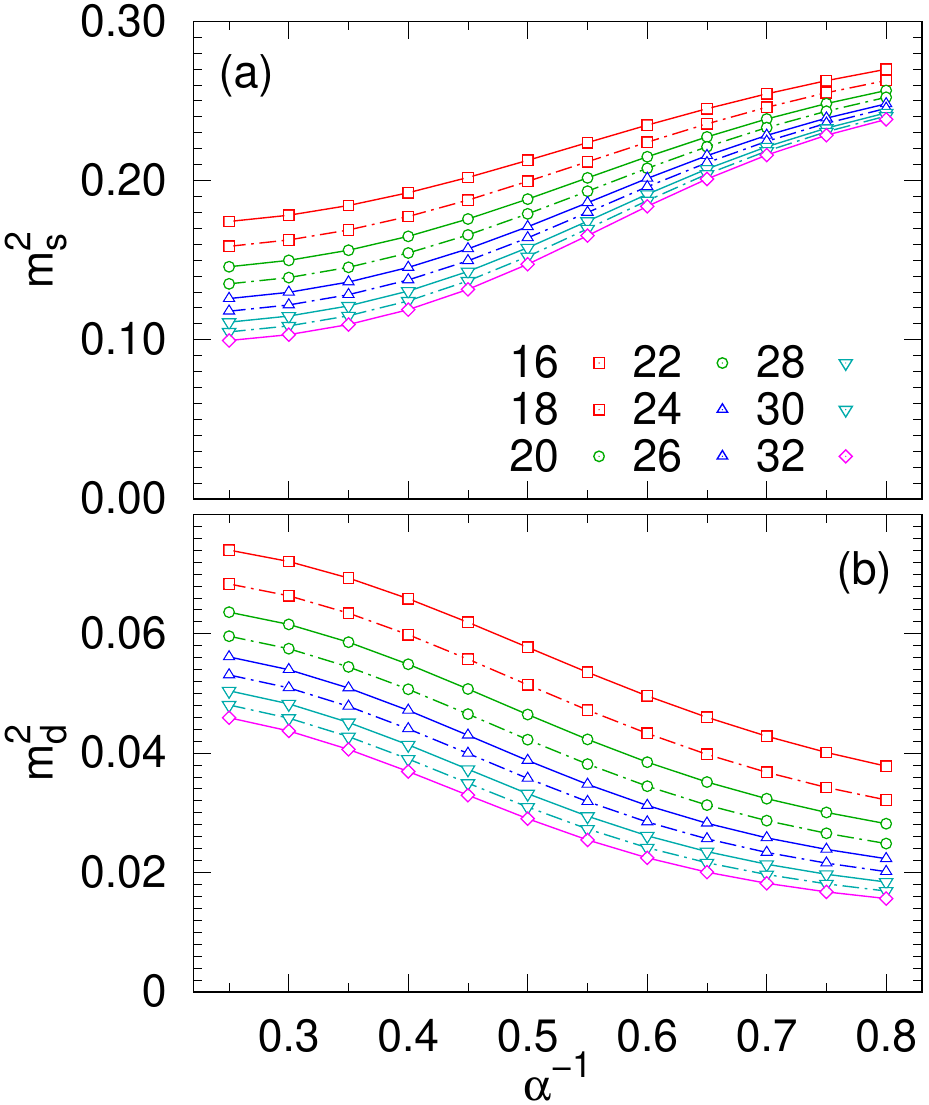}
\vskip-1mm
\caption{Squared order parameters of the chain model vs the inverse of the long-range interaction exponent.
(a) Staggered magnetization, Eq.~(\ref{msq}), and (b) dimer order, Eq.~(\ref{mdsq}).}
\label{orderchain}
\end{figure}

Results for the chain model are shown in Fig.~\ref{orderchain}. Here the trends versus the long-range parameter
$\alpha^{-1}$ are clearer than in the SSM, but when contrasting the two sets of results it should be kept in mind that
the system length is $N$ in the chain but $\sqrt{N}$ in the SSM, and the range of $N$ is similar in both cases. 
The length is of course what sets the cut-off for the correlation
functions and needs to be taken large to reach the asymptotic forms of the squared order parameters. Even in the 1D case, it is not 
possible to reliably extract the boundaries of the critical phase using the order parameters. This problem is well known from studies 
of the dimerization transition in the simpler $J_1$-$J_2$ chain and was a strong impetus for the development and use of the level 
crossing method \cite{Nomura92,Eggert96,Sandvik10ed}.

{\bf Conclusions and Discussion}---Our work presented here contributes to a growing sense that level spectroscopy is one of the most powerful generic methods
for detecting quantum phase transitions, not only in the well known context of 1D models \cite{Nomura92,Eggert96,Sandvik10,Nakamura00,Suwa15}
but also in 2D systems \cite{Lecheminant97,Misguich07,Suwa16,Schuler16,Wang18,Ferrari20,Nomura20,Yang22}. The main problem in 2D is that the
accessible system sizes are typically small, and finite-size extrapolations of level crossings---the aspect of the level spectrum on which we have
focused here---can be challenging. It is therefore important to extract the best possible information from the accessible level spectrum.
The main conclusion on methods to draw from our study is that relative gaps between two excited states (composite excitation gaps)
are useful alongside the conventional gaps relative to the ground state. In particular, the overall finite-size corrections of composite gap crossing
points can be smaller. We have further shown that cluster-shape effects can be significantly removed by consider relative distances between crossing
points of different gap combinations for the same system size. 

The power of the level crossing method to detect quantum phase transitions is further demonstrated by the very weak signals of the different
phases in the order parameters. Though the order parameters computed on larger cylinders \cite{Yang22} fully support the level-crossing values
of the phase boundaries, on the very small lattices accessible to Lanczos diagonalization they are not yet in the asymptotic regime where the 
size dependence can be reliably analyzed.

The primary model studied here, the SSM, is one of the key models of quantum magnetism, yet its QSL phase situated between the PSS and AFM phases
was only proposed very recently \cite{Yang22}. Our new results for the phase boundaries presented here are in remarkable agreement with the previous
results, considering the very small clusters used, $N \le 40$, while in Ref.~\cite{Yang22} different gap crossings of systems with up to $N=288$ spins
were studied. Furthermore, the boundary conditions are different, fully periodic here versus cylindrical in Ref.~\cite{Yang22}. This excellent agreement
between these two different calculations suggests that the empirically found finite-size scaling behavior of gap crossing points, with leading $1/N$ corrections,
is very robust and difficult to explain without the QSL phase located between the PSS and AFM phases.

For comparison, we also studied a Heisenberg chain with long-range unfrustrated interactions and short-range frustration. The initial goal of
this benchmark test was to investigate the same type of level crossings used for the SSM in a system where a QSL phase located between a two-fold
degenerate singlet ground state and an AFM phase is not controversial. Beyond the technical confirmation of the method, including reduced finite-size
corrections when a suitable composite gap is used instead of the conventional gaps relative to the ground state, the results also contain
useful information pertaining to the nature of the QSL phase in the SSM. Very similar size dependence is apparent of all the crossing points of the 1D
model in Fig.~\ref{criticalpoints2} and the analogous points for the SSM in Fig.~\ref{criticalpoints}, in particular the ordering of the crossing points
obtained with the different crossing gaps. This correspondence of low-energy states indicates that the two QSL phases (and critical points) have similar
deconfined spinon excitations despite the different dimensionalities, thus, they may have related field theory descriptions. We note that field theories
of 2D gapless spin liquids is an active field of investigation \cite{Hermele05,Chen17,Boyack18,Dupuis21,Liu21,Shackleton22}, and the spectral information
obtained here may help to determine the exact nature of the QSL phase in the SSM.

First-order direct PSS--AFM transitions have been studied in related 2D quantum spin models \cite{Zhao19} and were previously expected also in the
case of the SSM \cite{Corboz13,Xi21}. The most likely generic scenario for these 2D systems is a line of first-order transitions terminating at a
multi-critical deconfined
quantum-critical point, after which the QSL phase opens \cite{Yang22,Lu21}. A given model may then either undergo a first-order PSS--AFM transition
or cross the QSL phase, as we have argued here in the case of the SSM. We are currently exploring extended SSMs to further explore this scenario.

Recent NMR experiments on SCBO, have realized a PSS--AFM transition at low temperature, below $0.1$ K, driven by the strength of an external magnetic
field at high hydrostatic pressure \cite{Cui22}. The low transition temperature and observed scaling behaviors suggest a nearby critical point at field
strength close to $6$ T at a pressure slightly above $2.4$ GPa (the highest pressure studied). This critical point could possibly be the deconfined quantum
critical point terminating the SSM QSL phase at a finite magnetic field \cite{Yang22}, thus motivating further studies of the QSL
phase of the SSM with a magnetic field added to the Hamiltonian Eq.~(\ref{ssm}).

The SSM QSL phase at zero field may possibly be realized in SCBO somewhere between pressures of $2.6$
and $3.2$ GPa, where heat capacity measurements \cite{Guo20,Jimenez20} have not detected any phase transitions as a function of the temperature.
In SCBO, a complicating factor is that the weak inter-layer couplings  also will play some role \cite{Guo20,Sun21}, especially when perturbing a 2D gapless
phase. It would therefore also be important to study weakly coupled SSM layers.

\begin{acknowledgments}
{\it Acknowledgments}.--- L.W. was supported by the National Natural
Science Foundation of China, Grants No.~NSFC-11874080 and
No.~NSFC-11734002.  A.W.S. was supported by the Simons Foundation
(Simons Investigator Grant No.~511064).
\end{acknowledgments}


\begin{thebibliography}{99}

\bibitem{Balents05}
Balents, L. Spin liquids in frustrated magnets. Nature {\bf 464}, 199 (2010).

\bibitem{White11}  
S. Yan, D. A. Huse, and S. R. White, Spin-Liquid Ground State of the S = 1/2 Kagome Heisenberg Antiferromagnet, Science {\bf 332}, 1173 (2011).

\bibitem{Liao17}    
Gapless Spin-Liquid Ground State in the  Kagome Antiferromagnet
H. J. Liao, Z. Y. Xie, J. Chen, Z. Y. Liu, H. D Xie, R. Z Huang, B. Normand, and T. Xiang.
Phys. Rev. Lett. {\bf 118} 137202 (2017).

\bibitem{He17}
Y.-C. He, M. P. Zaletel, M. Oshikawa, and F. Pollmann,
Signatures of Dirac Cones in a DMRG Study of the Kagome Heisenberg Model,
Phys. Rev. X {\bf 7}, 031020 (2017).

\bibitem{Kitaev06}
A. Kitaev, Anyons in an exactly solved model and beyond, Ann. Phys. {\bf 321}, 2–111 (2006).
  
\bibitem{Nocera05}
M. P. Shores, E. A. Nytko, B. M. Bartlett, and D. G. Nocera,
A Structurally Perfect S = 1/2 Kagome Antiferromagnet,
J. Am. Chem. Soc. {\bf 127}, 13462 (2005).

\bibitem{Lee07}
 J. S. Helton {\it et al.},
Spin Dynamics of the Spin-1/2 Kagome Lattice Antiferromagnet ZnCu$_3$(OH)$_6$Cl$_2$, Phys. Rev. Lett. {\bf 98}, 107204 (2007).

\bibitem{Khaliullin10}
J. Chaloupka, G. Jackeli, and G. Khaliullin, Kitaev-Heisenberg model on a honeycomb lattice: possible exotic phases in iridium oxides A$_2$IrO$_3$,
Phys. Rev. Lett. {\bf 105}, 027204 (2010).

\bibitem{Norman16}
M. R. Norman,
Colloquium: Herbertsmithite and the search for the quantum spin liquid,
Rev. Mod. Phys. {\bf 88}, 041002 (2016).

\bibitem{Khuntia20}
P. Khuntia, M. Velazquez, Q. Barth\'elemy, F. Bert, E. Kermarrec, A. Legros, B. Bernu, L. Messio, A. Zorko, and P. Mendels,
Nat. Phys. {\bf 16}, 469 (2020).
  
\bibitem{Zheng17}
J. Zheng, K. Ran, T. Li, J. Wang, P. Wang, B. Liu, Z. X. Liu, B. Normand, J. Wen, W. Yu, 
Gapless Spin Excitations in the Field-Induced Quantum Spin Liquid Phase of $\alpha$-RuCl3,
Phys. Rev. Lett. {\bf 119}, 227208 (2017).

\bibitem{Li21}
H. Li {\it et al.},
Giant phonon anomalies in the proximate Kitaev quantum spin liquid $\alpha$-RuCl3, Nat. Commun. {\bf 2}, 3513 (2021).

\bibitem{Kimchi18a}
I. Kimchi, A. Nahum, and T. Senthil,
Valence Bonds in Random Quantum Magnets: Theory and Application to YbMgGaO$_4$, Phys. Rev. X {\bf 8}, 031028 (2018).

\bibitem{Liu18}
L. Liu, H. Shao, Y.-C. Lin, W. Guo, and A. W. Sandvik,
Random-Singlet Phase in Disordered Two-Dimensional Quantum Magnets
Phys. Rev. X {\bf 8}, 041040 (2018).

\bibitem{Kawamura19}
H. Kawamura and K. Uematsu, Nature of the randomness-induced quantum spin liquids in two dimensions,
J. Phys.: Condens. Matter {\bf 31}, 504003 (2019).

\bibitem{Li15}
Y. Li, G. Chen, W. Tong, L. Pi, J. Liu, Z. Yang, X. Wang, and Q. Zhang,
Rare-Earth Triangular Lattice Spin Liquid: A Single- Crystal Study of YbMgGaO$_4$, Phys. Rev. Lett. {\bf 115}, 167203 (2015).

\bibitem{Ma15}
Z. Ma et al., Spin-Glass Ground State in a Triangular-Lattice Compound YbZnGaO$_4$, Phys. Rev. Lett. {\bf 120}, 087201 (2018).

\bibitem{Kimchi18b}
I. Kimchi, J. P. Sheckelton, T. M. McQueen, and P. A. Lee,
Scaling and data collapse from local moments in frustrated disordered quantum spin systems,
Nat. Commun. {\bf 9}, 4367 (2018).

\bibitem{Kageyama99}
H. Kageyama, K. Yoshimura, R. Stern, N. V. Mushnikov, K. Onizuka, M. Kato, K. Kosuge, C. P. Slichter, T. Goto, and Y. Ueda, Exact Dimer Ground State and Quantized Magnetization Plateaus in the Two-Dimensional Spin System SrCu$_2$(BO$_3$)$_2$, Phys. Rev. Lett. {\bf 82}, 3168 (1999).

\bibitem{Waki07}
T. Waki, K. Arai, M. Takigawa, Y. Saiga, Y. Uwatoko, H. Kageyama, and Y. Ueda,
A novel ordered phase in SrCu$_2$(BO$_3$)$_2$ under high pressure,
J. Phys. Soc. Jpn. {\bf 76}, 073710 (2007).

\bibitem{Radtke15}
 G. Radtke, A. Saul, H. A. Dabkowska, M. B. Salamon, and M. Jaime, 
Magnetic nanopantograph in the SrCu$_2$(BO$_3$)$_2$ Shastry–Sutherland lattice, Proc. Natl. Acad. Sci. {\bf 112}, 1971 (2015).

\bibitem{Haravifard16}
S. Haravifard {\it et al.},
Crystallization of spin superlattices with pressure and field in the layered magnet SrCu$_2$(BO$_3$)$_2$,
Nat. Commun. {\bf 7}, 11956 (2016). 

\bibitem{Zayed17}
M. Zayed {\it et al.},
4-spin plaquette singlet state in the Shastry-Sutherland compound SrCu$_2$(BO$_3$)$_2$,
Nat. Phys. {\bf 13}, 962 (2017).

\bibitem{Bettler20}
S. Bettler, L. Stoppel, Z. Yan, S. Gvasaliya, and A. Zheludev, 
Sign switching of dimer correlations in SrCu$_2$(BO$_3$)$_2$ under hydrostatic pressure, 
Phys. Rev. Research {\bf 2}, 012010(R) (2020).

\bibitem{Guo20}
J. Guo, G. Sun, B. Zhao, L. Wang, W. Hong, V. A. Sidorov, N. Ma, Q. Wu, S. Li, Z. Y. Meng, A. W. Sandvik, and L. Sun, 
Quantum Phases of SrCu$_2$(BO$_3$)$_2$ from High-Pressure Thermodynamics,
Phys. Rev. Lett. {\bf 124}, 206602 (2020).

\bibitem{Jimenez20}
J. Larrea Jim\'enez {\it et al.},
A quantum magnetic analogue to the critical point of water, Nature {\bf 592}, 370 (2021).

\bibitem{Shastry81}
B. S. Shastry and B. Sutherland, 
Exact ground state of a quantum mechanical antiferromagnet,
Physica B+C {\bf 108}, 1069 (1981).

\bibitem{Lee19}
J. Y. Lee, Y.-Z. You, S. Sachdev, and A. Vishwanath,
Signatures of a Deconfined Phase Transition on the Shastry-Sutherland Lattice: Applications to
Quantum Critical SrCu$_2$(BO$_3$)$_2$, Phys. Rev. X {\bf 9}, 041037 (2019).

\bibitem{Sun21}
G. Sun, N. Ma, B. Zhao, A. W. Sandvik, and Z. Y. Meng,
Emergent O(4) symmetry at the phase transition from plaquette-singlet to antiferromagnetic order in quasi-two-dimensional quantum magnets,
Chin. Phys. B {\bf 30}, 067505 (2021).

\bibitem{Zhao19}
B. Zhao, P. Weinberg, and A. W. Sandvik,
Symmetry enhanced first-order phase transition in a two-dimensional quantum magnet,
Nature Phys. {\bf 15}, 678 (2019).

\bibitem{Yang22}
J. Yang, A. W. Sandvik, and L. Wang,
Quantum criticality and spin liquid phase in the Shastry-Sutherland model, Phys. Rev. B {\bf 105}, L060409 (2022).

\bibitem{Keles22}
A. Keles and E. Zhao,
Rise and fall of plaquette order in the Shastry-Sutherland magnet revealed by pseudofermion functional renormalization group,
Phys. Rev. B {\bf 105}, L041115 (2022).

\bibitem{Wang18}
L. Wang and A. W. Sandvik,
Critical Level Crossings and Gapless Spin Liquid in the Square-Lattice Spin-$1/2$ $J_1$-$J_2$ Heisenberg Antiferromagnet,
Phys. Rev. Lett. {\bf 121}, 107202 (2018).

\bibitem{Gong14}
S.-S. Gong, W. Zhu, D. N. Sheng, O. I. Motrunich, and M. P. A. Fisher, 
Plaquette Ordered Phase and Quantum Phase Diagram in the Spin-$1/2$ $J_1$-$J_2$  Square Heisenberg Model, 
Phys. Rev. Lett. {\bf 113}, 027201 (2014).

\bibitem{Morita15}
S. Morita, R. Kaneko, and M. Imada, 
Quantum spin liquid in spin $1/2$ $J_1$-$J_2$ Heisenberg model on square lattice: Many-variable variational Monte Carlo study combined 
with quantum-number projections, 
J. Phys. Soc. Jpn. {\bf 84}, 024720 (2015).

\bibitem{Ferrari20}
F. Ferrari and F. Becca, 
Gapless spin liquid and valence-bond solid in the $J_1$-$J_2$ Heisenberg model on the square lattice: 
Insights from singlet and triplet excitations,
Phys. Rev. B {\bf 102}, 014417 (2020).

\bibitem{Nomura20}
Y. Nomura and M. Imada,
Dirac-type nodal spin liquid revealed by machine learning,
Phys. Rev. X {\bf 11}, 031034 (2021).

\bibitem{Schackleton21}
H. Shackleton, A. Thomson, and S. Sachdev,
Deconfined criticality and a gapless $\mathbb{Z}_2$ spin liquid in the square lattice antiferromagnet,
Phys. Rev. B {\bf 104}, 045110 (2021).

\bibitem{Laflorencie04}
N. Laflorencie and D. Poilblanc,
Simulations of pure and doped low-dimensional spin-$1/2$ gapped systems,
Lecture Notes in Physics {\bf 645}, 227 (2004).

\bibitem{Noack05}
R.M. Noack and S. Manmana, Diagonalization and Numerical Renormalization-Group-Based Methods for Interacting Quantum Systems,
AIP Conf. Proc. {\bf 789}, 93 (2005).
  
\bibitem{Weisse08}
A. Weisse, H. Fehske, Exact Diagonalization Techniques, Lecture Notes in Physics {\bf 739}, 529 (2008).  
  
\bibitem{Lauchli11}
A. L\"auchli, "Numerical Simulations of Frustrated Systems" in
{\it Highly Frustrated Magnets}, edited by C. Lacroix, P. Mendels, and F. Mila, (Springer Verlag, 2011).
  
\bibitem{Sandvik10ed}
A. W. Sandvik, Computational Studies of Quantum Spin Systems, AIP Conf. Proc. {\bf 1297}, 135 (2010). 

\bibitem{Nomura92}
K. Nomura and K. Okamoto, 
Fluid-dimer critical point in $S=1/2$ antiferromagnetic Heisenberg chain with next nearest neighbor interactions, 
Phys. Lett. A {\bf 169}, 433 (1992).

\bibitem{Eggert96}
S. Eggert, 
Numerical evidence for multiplicative logarithmic corrections from marginal operators, 
Phys. Rev. B {\bf 54}, R9612 (1996).

\bibitem{Sandvik10}
A. W. Sandvik, 
Ground States of a Frustrated Quantum Spin Chain with Long-Range Interactions, 
Phys. Rev. Lett. {\bf 104}, 137204 (2010).

\bibitem{Suwa16}
H. Suwa, A. Sen, and A. W. Sandvik, 
Level spectroscopy in a two-dimensional quantum magnet: Linearly dispersing spinons at the deconfined quantum critical point, 
Phys. Rev. B {\bf 94}, 144416 (2016).

\bibitem{Laflorencie05}
N. Laflorencie, I. Affleck, and M. Berciu,
Critical phenomena and quantum phase transition in long range Heisenberg antiferromagnetic chains,
J. Stat. Mech. (2005) P12001.

\bibitem{koga00}
A. Koga and N. Kawakami,
Quantum Phase Transitions in the Shastry-Sutherland Model for SrCu$_2$(BO$_3$)$_2$,
Phys. Rev. Lett. {\bf 84}, 4461 (2000).

\bibitem{Corboz13}
P. Corboz and F. Mila,
Tensor network study of the Shastry-Sutherland model in zero magnetic field,
Phys. Rev. B {\bf 87}, 115144 (2013).

\bibitem{Anderson59}
P. W. Anderson, New Approach to the Theory of Superexchange Interactions,
Phys. Rev. B {\bf 115}, 2 (1959).

\bibitem{dmrgnote}
Though these triplets with energy between $E(T_1)$ and $E(T_2)$ have lattice quantum numbers
different from those of $T_1$ and $T_2$, the corresponding symmetries are not implemented in the DMRG
calculation (but the total spin symmetry is implemented in this case), only computed as expectation values
with the states obtained. The states have to be generated one-by-one starting from the lowest one, and
convergence of this procedure becomes increasingly challenging with the number of states computed
\cite{Wang18,Yang22}, and we have not been able to reach the the state with quantum numbers corresponding to $T_2$ for $N=40$.

\bibitem{Nakamura00}
M. Nakamura, Tricritical behavior in the extended Hubbard chains,
Phys. Rev. B {\bf 61}, 16377 (2000).

\bibitem{Suwa15}
H. Suwa and S. Todo,
Generalized Moment Method for Gap Estimation and Quantum Monte Carlo Level Spectroscopy,
Phys. Rev. Lett. {\bf 115}, 080601 (2015). 

\bibitem{Lecheminant97}
P. Lecheminant, B. Bernu, C. Lhuillier, L. Pierre, and P. Sindzingre,
Order versus disorder in the quantum Heisenberg antiferromagnet on the kagomé lattice using exact spectra analysis,
Phys. Rev. B {\bf 56}, 2521 (1997).

\bibitem{Misguich07}
G Misguich and P Sindzingre,
Detecting spontaneous symmetry breaking in finite-size spectra of frustrated quantum antiferromagnets,
J. Phys.: Condens Matter {\bf 19}, 145202 (2007).

\bibitem{Schuler16}
M. Schuler, S. Whitsitt, L. P. Henry, S. Sachdev, and A. M. L\"auchli, Universal Signatures of Quantum Critical Points from
Finite-Size Torus Spectra: A Window into the Operator Content of Higher-Dimensional Conformal Field Theories,
Phys. Rev. Lett. {\bf 117}, 210401 (2016).

\bibitem{Hermele05}
M. Hermele, T. Senthil, and M. P. A. Fisher, Algebraic spin liquid as the mother of many competing orders,
Phys. Rev. B {\bf 72}, 104404 (2005).

\bibitem{Chen17}
J.-Yao Chen and D. Poilblanc,
Topological $Z_2$ resonating-valence-bond spin liquid on the square lattice,
Phys. Rev. B {\bf 97}, 161107(R) (2017).

\bibitem{Boyack18}
R. Boyack, C.-H. Lin, N. Zerf, A. Rayyan, and J. Maciejko, Transition between algebraic and $F_2$
quantum spin liquids at large n,” Phys. Rev. B {\bf 98}, 035137 (2018).

\bibitem{Dupuis21}
E. Dupuis, R. Boyack, and W. Witczak-Krempa, Anomalous dimensions of monopole operators at the transitions
between Dirac and topological spin liquids, arXiv:2108.05922.

\bibitem{Liu21}
W.-Y. Liu, J. Hasik, S.-S. Gong, D. Poilblanc, W.-Q. Chen, and Z.-C. Gu,
The emergence of gapless quantum spin liquid from deconfined quantum critical point,
arXiv:2110.11138.

\bibitem{Shackleton22}
H. Shackleton and S. Sachdev, Anisotropic deconfined criticality in Dirac spin liquids,
arXiv:2203.01962.  
  
\bibitem{Xi21}
N. Xi, H. Chen, Z. Y. Xie, and R. Yu,
First-order transition between the plaquette valence bond solid and antiferromagnetic phases of the Shastry-Sutherland model,
arXiv:2111.07368.

\bibitem{Lu21}
D.-C. Lu, C. Xu, and Y.-Z. You,
Self-duality protected multi-criticality in deconfined quantum phase transitions,
Phys. Rev. B {\bf 104}, 205142 (2021).

\bibitem{Cui22}
Y. Cui, L. Liu, H. Lin, K.-H. Wu, W. Hong, X. Liu, C. Li, Z. Hu, N. Xi, S. Li, R. Yu, A. W. Sandvik, W. Yu,
Deconfined quantum criticality and emergent symmetry in SrCu$_2$(BO$_3$)$_2$, arXiv:2204.08133.


\end{thebibliography}
\end{document}